\documentclass[aps,prb,twocolumn,
	groupedaddress,superscriptaddress,
	amsfonts,amssymb,amsmath,floatfix,
	citeautoscript]{revtex4-1}
\usepackage{graphicx} 
\usepackage[centering,hmargin=22mm,tmargin=26mm,bmargin=26mm]{geometry}
\usepackage{amsmath}
\usepackage{hyperref} 
\usepackage{multirow}
\usepackage{newtxtext}
\usepackage{booktabs}
\usepackage{xcolor} 
\usepackage{hyperref}
\usepackage[normalem]{ulem}
\usepackage[textwidth=1.5cm,textsize=footnotesize]{todonotes}
\hypersetup{colorlinks, 
	linkcolor={blue!75!black!80!yellow},
	citecolor={blue!75!black!80!yellow}, 
	urlcolor={blue!75!black!80!yellow}
	}
\usepackage[cmintegrals]{newtxmath}
\makeatletter
\renewcommand\@make@capt@title[2]{	\@ifx@empty\float@link{\@firstofone}{\expandafter\href\expandafter{\float@link}}	\sffamily{\textbf{#1}}\@caption@fignum@sep#2
}
\makeatother

\newcommand{\HarvardSEAS}{John A. Paulson School of Engineering and Applied
Sciences, Harvard University, Cambridge, MA, USA}
\newcommand{\HarvardCCB}{Department of Chemistry and Chemical Biology, Harvard
University, Cambridge, MA, USA} 
\newcommand{\OxfordMS}{Department of Materials, University of Oxford, Oxford, Oxfordshire, UK}

\newcommand{\two}{$2\times2\times2$ }

\newcommand{\sub}[1]{\ensuremath{_{\textrm{#1}}}} 

\begin{document}
\author{Yuxin Yin}\affiliation{\HarvardSEAS}\affiliation{\OxfordMS}
\author{Jennifer Coulter}\affiliation{\HarvardSEAS}
\author{Christopher J. Ciccarino }\affiliation{\HarvardSEAS}\affiliation{\HarvardCCB}
\author{Prineha Narang}\email{prineha@seas.harvard.edu}\affiliation{\HarvardSEAS}

\title{A Theoretical Investigation of Charge Density Wave Instability in CuS\sub{2}}

\date{\today}

\begin{abstract} 
The existence of a charge density wave (CDW) in transition metal dichalcogenide CuS\sub{2} has remained undetermined since its first experimental synthesis nearly 50 years ago. Despite conflicting experimental literature regarding its low temperature structure, there exists no theoretical study of the phonon properties and lattice stability of this material.
By studying the first-principles electronic structure and phonon properties of CuS\sub{2} at various electronic temperatures, we identify temperature-sensitive soft phonon modes which unveil a previously unreported Kohn anomaly at approximately 100K. Variation of the electronic temperature shows the presence of two distinct phases, characterized at low temperature by a \two periodic charge modulation associated with the motion of the S\sub{2} dimers.
Investigation of the Fermi surface presents a potential Fermi surface nesting vector related to the location of the Kohn anomaly and observed band splittings in the unfolded bandstructure. 
The combination of these results suggests a strong possibility of CDW order in CuS\sub{2}. Further study of CuS\sub{2} in monolayer form finds no evidence of a CDW phase, as the identified bulk periodic distortions cannot be realized in 2D. This behavior sets this material apart from other transition metal dichalcogenide materials, which exhibit a charge density wave phase down to the 2D limit. 
As CDW in TMDC materials is considered to compete with superconductivity, the lack of CDW in monolayer CuS\sub{2} suggests the possibility of enhanced superconductivity relative to other transition metal dichalcogenides. Overall, our work identifies CuS\sub{2} as a previously unrealized candidate to study interplay of superconductivity, CDW order, and dimensionality.
\end{abstract}

\maketitle
Transition metal dichalcogenides (TMDCs) like Ta, 
Nb(S,Se)\sub{2} have displayed interesting physics including charge density wave (CDW) formation and low temperature superconductivity~\cite{castro_neto_charge_2001,rossnagel_origin_2011,cudazzo_plasmon_2012,PhysRevB.97.035111,Basov:2017,keimer2017physics,cho_using_2018}. While bulk, layered TMDC structures can host CDW, recent studies have investigated the competition between CDW order and superconductivity,~\cite{chen_charge_2015,chang_direct_2012,cho_using_2018} as well as the enhancement of superconducting T\sub{c} in monolayer TMDCs in relation to the suppression or lack of CDW order~\cite{PhysRevLett.28.299,yang_enhanced_2018,wei_manipulating_2017,navarro2016enhanced,xi2015strongly,kang2011suppression}. 
Despite a lack of a unified understanding of the origins of CDW in TMDCs, thus far, studied examples have displayed similar CDW characteristics down to the 2D limit, possibly limiting T\sub{c} enhancement. 

Though the superconducting behavior of TMDC CuS\sub{2} is well known~\cite{bither_transition_1968}, there exists a long-standing debate about the existence of CDW order in this material~\cite{krill_magnetic_1976, ueda_copper_2002}. 
Initial studies of bulk susceptibility~\cite{bither_transition_1968, gautier_magnetic_1974}, nuclear magnetic resonance~\cite{gautier_magnetic_1974, krill_magnetic_1976,kontani_specific_2000}, and specific heat~\cite{krill_magnetic_1976, kontani_specific_2000} have presented evidence of a second order structural phase transition at 150K, hinting at the formation of a CDW structure. However, a more recent experiment rejected this possibility as a result of a contradictory Hall coefficient measurement~\cite{ueda_copper_2002}, and this stance has been adopted by subsequent studies.

Throughout this experimental debate, theoretical study of CuS\sub{2} has remained extremely limited~\cite{bullett_electronic_1982, temmerman_electronic_1993, kakihana_superconducting_2018}. Presently, there are no reports of the phonon properties of CuS\sub{2}, which could provide critical insight regarding its structural phases. Theoretical studies of other CDW materials, including NbSe\sub{2}~\cite{calandra_effect_2009}, TiSe\sub{2}~\cite{duong_ab_2015,bianco_electronic_2015} and TaSe\sub{2}~\cite{yan_structural_2015}, have successfully captured CDW behavior and related phenomena through analysis of electronic-temperature dependent structural and electronic properties. 

In this context, we present \emph{ab initio} 
calculations of the electronic and phononic properties of CuS\sub{2} as a function of electronic temperature\cite{PhysRevB.94.075120, PhysRevB.97.195435}. We observe that the experimentally reported high-temperature structure~\cite{bither_transition_1968} exhibits instabilities and a Kohn anomaly in the phonon dispersion for calculations at low electronic temperature. By constructing and stabilizing a superstructure of CuS\sub{2}, we report for the first time the existence of a \two periodic manifestation of this material. We find that the transition to the distorted structure is governed by a twice-periodic displacement of the S\sub{2} dimers of CuS\sub{2}, resulting in a modulation of the charge density of the same periodicity. By evaluating the electronic structure of the low-temperature distorted structure, we observe the appearance of band splittings and identify a potential Fermi surface nesting vector associated with the Kohn anomaly. Subsequently, we investigate monolayer CuS\sub{2}~\cite{liu_observation_2018} and find no instabilities or any other hints of CDW formation, regardless of electronic temperature, possibly because it cannot host the dimer motions observed in the bulk structural distortion. 

Together, our results strongly suggest CDW order in bulk CuS\sub{2}, and by identifying its origins, explain the absence of CDW in the 2D limit. We therefore present it as an exemplary platform to study the competition between superconductivity and CDW order, and in particular, the role dimensionality plays in these effects. We discuss the unique aspects of CDW formation relative to well-known TMDC materials, and further discuss pathways for enhancing superconductivity in both bulk and monolayer CuS\sub{2}.

In textbook 1D systems, the origin of CDW is conventionally analyzed in the Peierls instability picture~\cite{peierls_zur_1930, peierls_1956, Frohlich_theory_1954} where the system undergoes a periodic lattice distortion characterized by charge density modulation to form a symmetry-reduced state. This results in band splittings at the Fermi wavevector ($\vec{k}\sub{F}$) related to a Fermi surface nesting vector $\vec{k}\sub{nesting}=2\vec{k}\sub{F}$, which connects multiple points on the Fermi surface and indicates the location of a Kohn anomaly in the phonon dispersion~\cite{kohn_image_1959}. 
However, in higher dimensional systems, the analogy to the Peierls instability is not as straightforward~\cite{johannes_fermi_2008} and Fermi surface nesting alone is generally not sufficient for understanding the origin of CDW formation. 
Other mechanisms include the excitonic insulator instability~\cite{bianco_electronic_2015, porer_non-thermal_2014}, Jahn-Teller effects~\cite{porer_non-thermal_2014}, momentum-dependent electron-phonon coupling~\cite{johannes_fermi_2008, zhu_classification_2015, calandra_effect_2009} as well as a combination of these mechanisms. 

Therefore, to investigate the occurrence of CDW in CuS\sub{2}, we begin by considering its structural stability. We start with the experimental pyrite (Pa3) structure measured at ambient temperatures~\cite{bither_transition_1968, krill_magnetic_1976},
with Cu atoms occupying
fcc sites of the unit cell, and S\sub{2} dimers centered along cell edges. After structural relaxation, we perform density 
functional theory (DFT)\cite{JDFTx} calculations with computational procedures previously introduced\cite{nanoLett:CC, Narang:2017, Brown:2016, NatCom}. To describe CuS\sub{2}, 
we selected the ultrasoft RRKJ pseudopotentials\cite{rappe_optimized_1990, dal_corso_pseudopotentials_2014} parameterized for the PBE exchange-correlation functional\cite{PBE}. We also include a Hubbard U 
parameter (DFT+U) of $U_{\mathrm{eff}}=0.5$ eV, as determined from first principles linear response~\cite{CococcioniDFT+U}. This relatively small $U_{\mathrm{eff}}$ is consistent with experimental studies which have found weak electronic correlation~\cite{ueda_copper_2002,kakihana_superconducting_2018} in CuS\sub{2}, as we find the bands near the Fermi energy are mostly of Sulfur $p$ (as opposed to Copper $d$) character.

In order to probe temperature-dependent phenomena in CuS\sub{2}, we use the Fermi-Dirac smearing scheme to mimic an 
electronic temperature. All calculations where the smearing width or 
electronic temperature is not specified use a 0.001 Hartree smearing (equivalent to an electronic temperature of 315K).
Finally, to compare the electronic structure of 
low- and high-symmetry structures, we apply a band unfolding technique.
Here we determine the overlap of the supercell Bloch states
with those of the primitive cell using a symmetry-averaged
spectral function technique~\cite{tian_correlating_2019}, as outlined in Ref.~\citenum{Medeiros:2014}. 

\begin{figure}[h!]
\includegraphics[width=1\columnwidth]{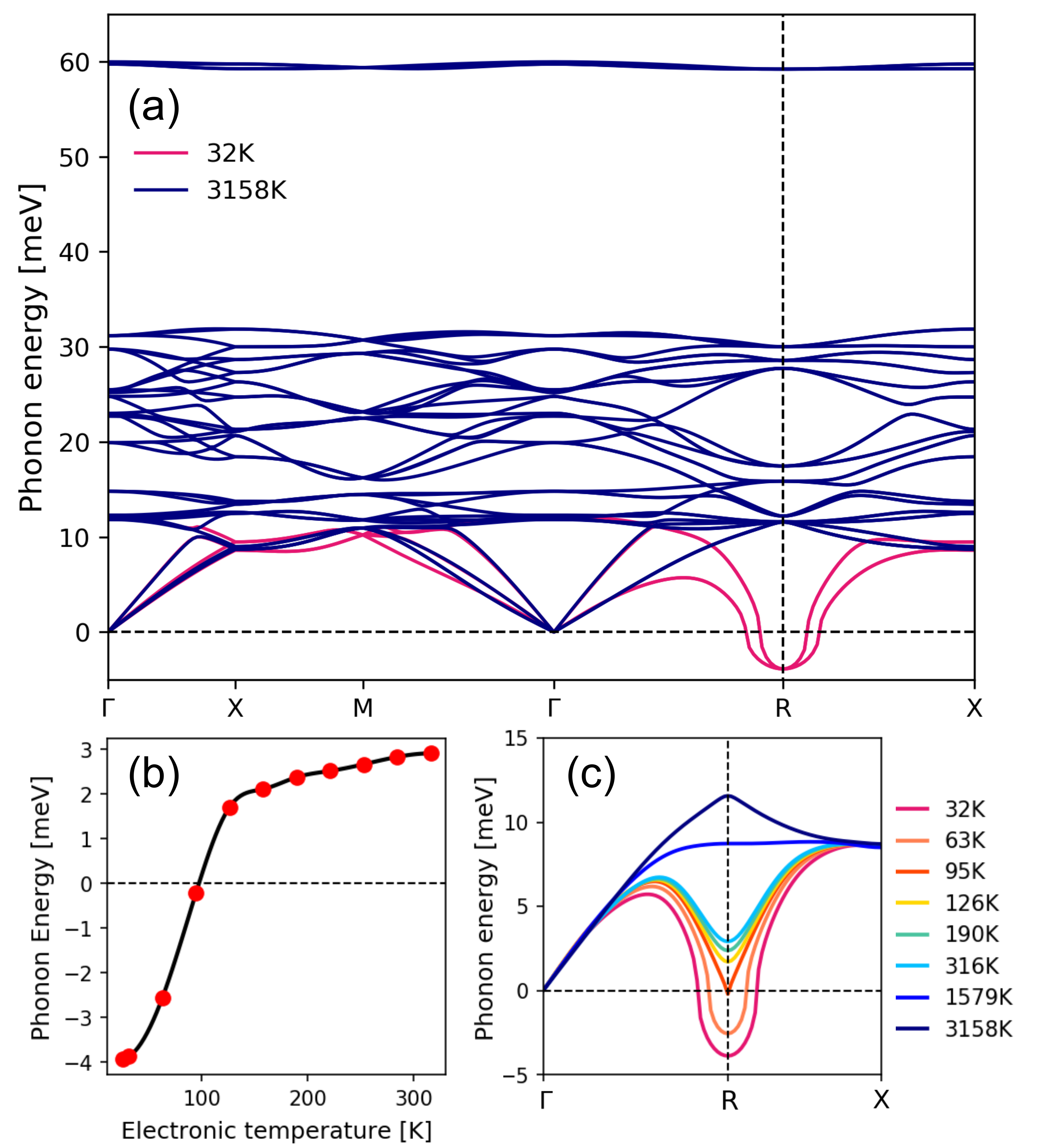} 
\caption{\textbf{Structural Stability of CuS\sub{2}}. The (a) predicted phonon dispersion for the CuS\sub{2} unit cell experimentally established in high temperature measurements \cite{bither_transition_1968} when calculated for low and high electronic temperatures along with (b) the lowest phonon frequency as a function of electronic temperature and (c) a close up of the instability showing its 
temperature dependence and the Kohn anomaly observed near 100K. From these calculations, we uncover a key difference between the low and high temperature structures of CuS\sub{2} which indicates the existence of a structural phase transition. } 
\label{fig:12atomCell}
\end{figure}

Initial calculations of the 
phonon dispersion of the experimental, high-temperature structure using low-temperature smearing reveal four soft 
phonon modes at the R point shown in Fig.~\ref{fig:12atomCell}(a). Upon varying the smearing parameter, we found these soft modes near this wavevector to be strongly sensitive to the electronic temperature, as shown in Fig.~\ref{fig:12atomCell}(b) and (c).
We note the appearance of a discontinuity in the phonon dispersion (a Kohn 
anomaly) located at wavevector $\vec{q}=\vec{R}$ for an electronic temperature corresponding to 
approximately 100K.

\begin{figure*}[!ht]\includegraphics[width=\textwidth]{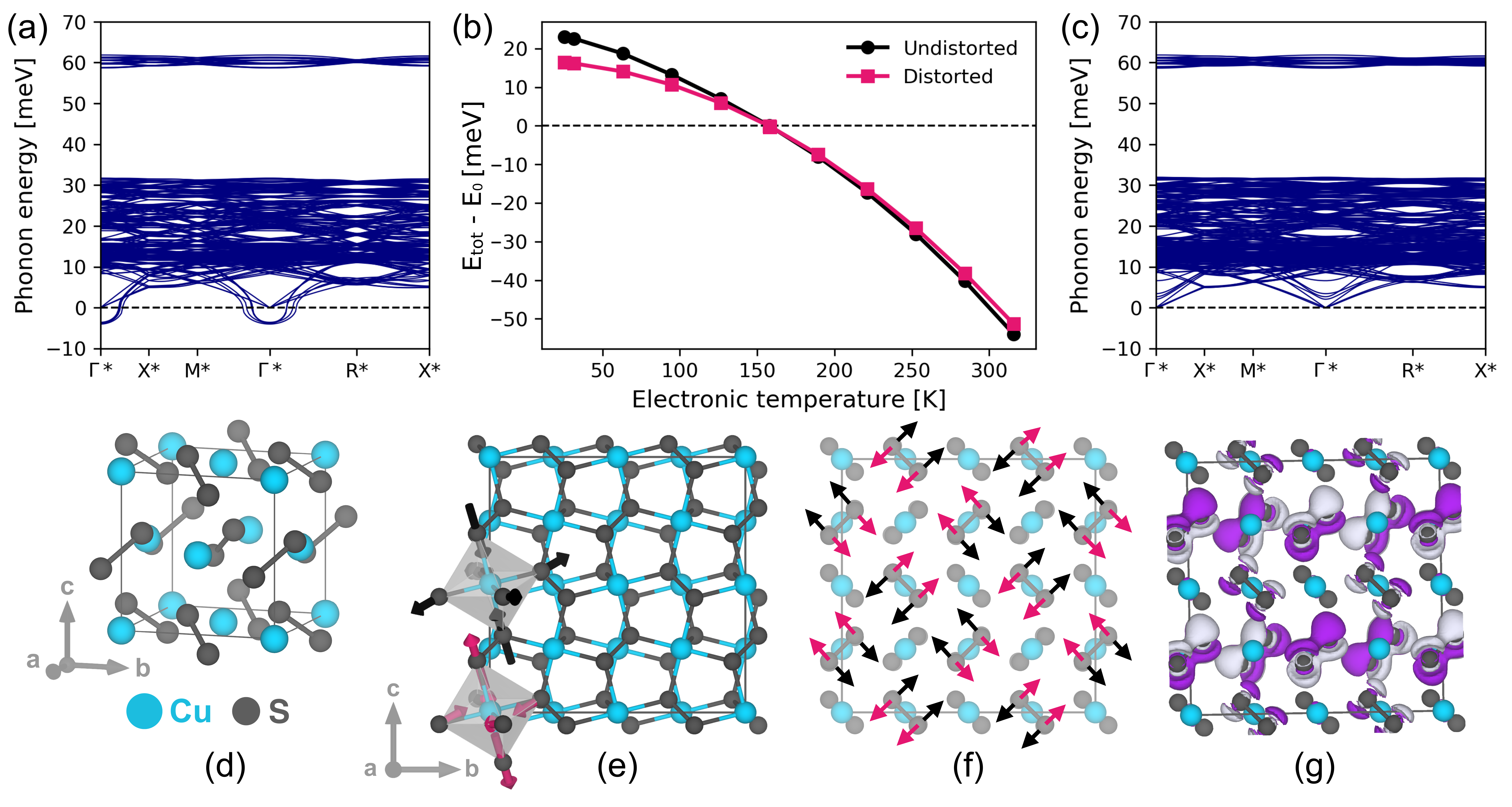}
\caption {\textbf{Collective Distortion and the Appearance of CDW Order in CuS\sub{2}}. (a) The predicted phonon dispersion for the undistorted superstructure (with * on the band path labeling high symmetry points in the supercell), alongside (b) the ground state energies for the two stabilized structures (relative to the energy of the undistorted structure at 150K) with respect to electronic temperature, showing an energy crossover and indicating a phase transition at 150K, in remarkable agreement with experimental measurements~\cite{krill_magnetic_1976}. In (c), we show the 
phonon dispersion of the distorted superstructure after applying the stabilizing collective mode depicted in (f), which creates the symmetry-reduced distorted structure and corresponding alternating octahedral 
compression and expansion shown in (e). By taking the difference in 
charge density between the undistorted and distorted structures, we 
find the periodic charge density modulation presented in (g), a strong 
indicator of CDW formation. The realization of this stable distorted superstructure at low electronic temperatures agrees well with early experimental reports\cite{vanderschaeve_electron_1976}, and suggests the possibility of CDW formation in CuS\sub{2}.} 
\label{fig:222supercell}
\end{figure*}

Motivated by past experimental work suggesting the existence of a \two 
superstructure in CuS\sub{2} at low temperatures 
\cite{vanderschaeve_electron_1976}, we sought to resolve the 
structural instability by using a \two supercell. 
The predicted phonon dispersion of the constructed  
supercell (depicted in Fig.~\ref{fig:222supercell}(a)) reveals four negative 
phonon modes at an electronic temperature of 32K, consistent with the instabilities observed in the unit cell structure. We note 
the soft modes observed are at the $\Gamma^*$ point of the supercell, which 
is equivalent to R point in the original cell due to the corresponding change in the Brillouin zone of the supercell structure (see Fig.~1 in SI). 

We identify the atomic displacements resulting from the four unstable 
phonon modes and observe each mode corresponds to \two periodic side-ways
motions of S\sub{2} dimers, which cannot be captured in the single 
unit cell. 
These four modes collectively manifest themselves as an effective motion of the S\sub{2} dimers with 
a period of $2a$, twice the original unit cell lattice coefficient, as schematically shown in Fig.~\ref{fig:222supercell}(f).

We perform energy minimization with respect to the ground state energy of the high-symmetry phase as a function of the magnitude of the eigendisplacements of each of these modes (for illustration of the individual modes and energy minimization, see 
Fig.~2 in SI).
By applying this collective distortion to the constructed undistorted
superstructure, we obtain a distorted structure with reduced symmetry, 
from (Pa3 to P-1) as shown in Fig.~\ref{fig:222supercell}(e), and predict a stable phonon dispersion at low electronic temperatures (seen in Fig.~\ref{fig:222supercell}(c)). These collective distortions lead to an alternating expansion and 
compression of Cu centered octahedra, as shown in Fig.~\ref{fig:222supercell}(d), which could introduce a crystal field splitting of the $d$ orbital energy levels and may impact electron occupation and charge distribution near the Cu sites. By comparing 
the undistorted and distorted structures and calculating the difference 
in their charge densities, we identify a 2$a$ periodic modulation of charge 
density, commensurate with the periodicity of the distorted structure, as shown in Fig.~\ref{fig:222supercell}(g).

Additionally, we investigate the relative stability of the undistorted and distorted 
supercells (Fig.\ref{fig:222supercell}(b)), and find that as we change electronic 
temperature of both structures, we see a crossover in the ground state energy 
at about 150K, a transition that would agree impressively well 
with experimental results~\cite{krill_magnetic_1976} and correspond to what was calculated in 
Fig.~\ref{fig:12atomCell}(b,c). When relaxed at higher electronic 
temperature, the distorted structure returns to the undistorted 
structure, indicating the process can be reversed.

The combination of 
the observed Kohn anomaly, periodic lattice distortion, periodic charge density modulation, and temperature-dependent phase crossover are all indicative of the presence of CDW in bulk CuS\sub{2}. Our results therefore capture, for the first time in a theoretical study, a stable low temperature distorted phase of CuS\sub{2} exhibiting strong indicators of CDW ordering.

\begin{figure}[h!]
\includegraphics[width=1\columnwidth]{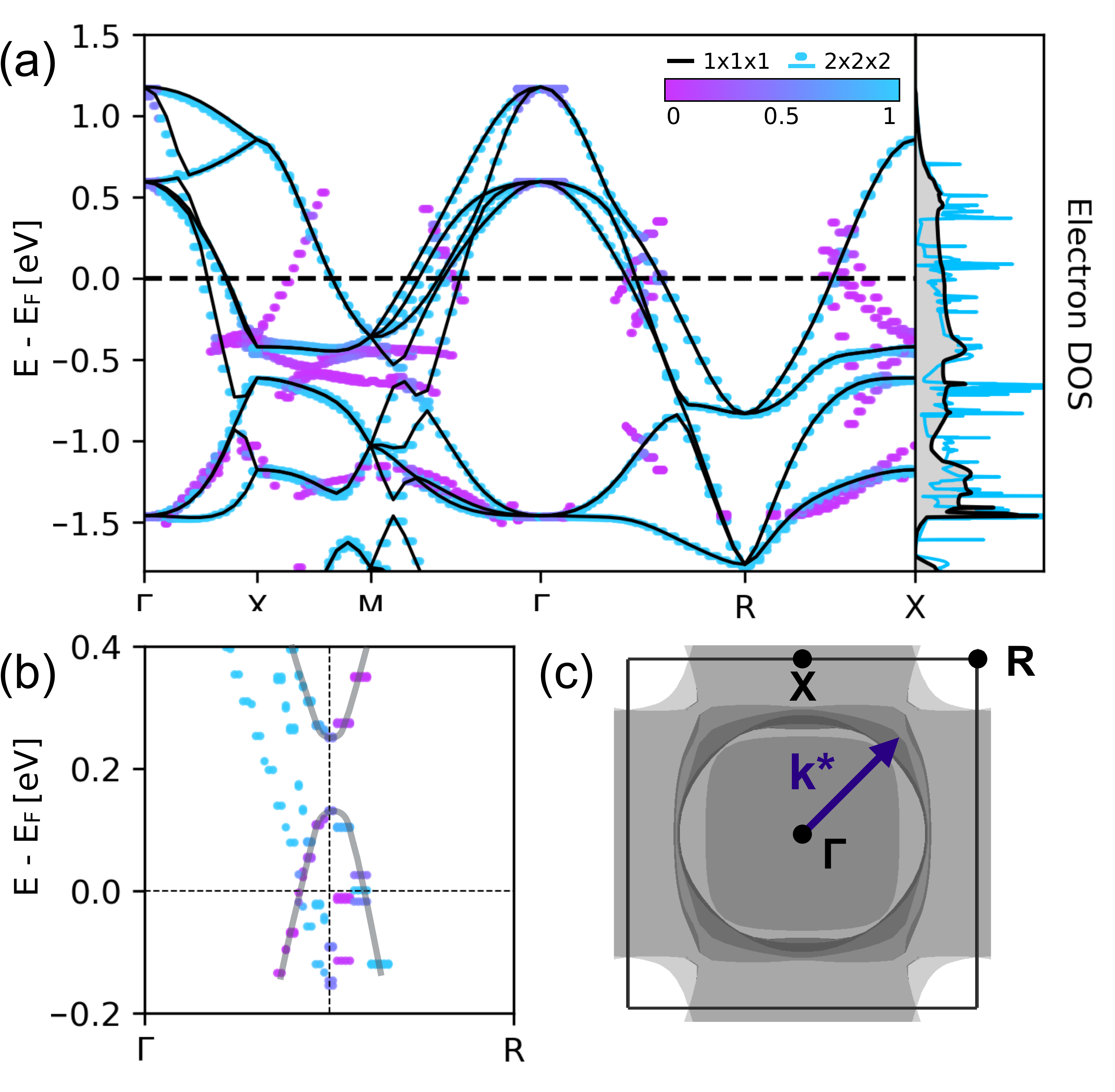}
\caption{\textbf{Band Splittings and Fermi Surface  Nesting}.
Panel (a) compares the unfolded bandstructure of the distorted phase ($2\times2\times2$) with 
that of the undistorted phase ($1\times1\times1$) superimposed as black lines. The color for the distorted phase 
indicates the spectral weight of each point, calculated via band unfolding. Also presented is the 
the density of states for the distorted structure (blue) and 
undistorted structure (black/grey). A zoom-in on the unfolded bandstructure in (b)
highlights (with grey guidelines) the splitting of the bands between the $\Gamma$ and R point.
(c) The Fermi surface of the 
undistorted phase, showing $\vec{k}^*$ halfway between $\Gamma$ and R,
which appears to connect a number of locations on the Fermi surface and indicates a link between Fermi surface nesting and CDW in CuS\sub{2}. }  
\label{fig:electronicOrigins}
\end{figure}

To understand the electronic origins of the potential CDW phase, we investigate the corresponding changes in the electronic structure.
By using a band unfolding technique, we compare the effective band structure of the distorted supercell with that of the undistorted structure, seen in 
Fig.~\ref{fig:electronicOrigins}(a). 
We observe that the distortion results in multiple band splittings and formation of small gaps at multiple points along the band path. The locations of these gaps tend to occur halfway between high symmetry points (e.g., $\Gamma$ and R, R and X) and slightly above the Fermi energy level, in contrast to gap openings observed at $E_{\mathrm{F}}$ in other CDW materials~\cite{yang_enhanced_2018,singh_stable_2017,rossnagel_origin_2011}.

We particularly focus on the splitting at approximately
0.2 eV above the Fermi level
(Fig.~\ref{fig:electronicOrigins}(b)), which is centered at a wavevector $\vec{k}^* = \frac{1}{2}\vec{R}$, half that of the vector where the Kohn anomaly was observed in the 
undistorted structure). We find that this wavevector, when overlaid with the Fermi surface shown in Fig.~\ref{fig:electronicOrigins}(c), points to a saddle point in the surface, and that twice this vector ($2\vec{k}^*$) connects such saddle points diagonally across the Fermi surface. 
In the conventional 1D model of CDW 
formation~\cite{Frohlich_theory_1954,peierls_1956}, the Peierls instability 
results in band gap formation at $\vec{k}\sub{F}$ and a Kohn anomaly located at $\vec{k}=2\vec{k}\sub{F}$. 
This appears to be the case in CuS\sub{2}, indicating that Fermi surface nesting could be responsible for CDW formation in this material, contrary what occurs in NbSe\sub{2}~\cite{weber_extended_2011,johannes_fermi_2008} and 
TaSe\sub{2}~\cite{johannes_fermi_2008} where no effective nesting exists at the CDW wavevector. 

Additionally, we find that CuS\sub{2} shows weak electronic correlation~\cite{sachdev_charge-_1995}
due to the dominance of sulfur $p$ rather than Cu $d$ orbitals at the 
Fermi level, while the gaps are energetically away from the Fermi level. This suggests a Mott-Hubbard-type metal-insulator transition~\cite{johnston_charge_2014} cannot provide a significant contribution to CDW formation.
We also note that the splitting of the $d$ orbital energy 
level as a result of compression and expansion of alternating Cu site
octahedral environments, which would account for the change in bandstructure approximately
0.5 eV below the Fermi energy level between the X and M points. 
This is supported by analyzing the relative
contribution of $e_g$ and $t_{2g}$ characteristics to the density of
states (see SI Fig.~3). The resulting occupation shifts in the $e_g$ orbitals are due to shifts in $d$ shells, which, because they are energetically positioned below the Fermi level, cannot strongly contribute to the understanding of the observed gaps located slightly above the Fermi energy.

Despite the hints at Fermi surface nesting related CDW and band gap formation in bulk CuS\sub{2}, we note that there are a number of possible mechanisms which could result in a gap opening~\cite{dessau_k_1998} and CDW formation, including 
the Jahn-Teller effect~\cite{porer_non-thermal_2014} and the electron-phonon 
interaction~\cite{wijayaratne_spectroscopic_2017,weber_extended_2011}, which require further study in CuS\sub{2}. 

\begin{figure}[!h]
\includegraphics[width=\columnwidth]{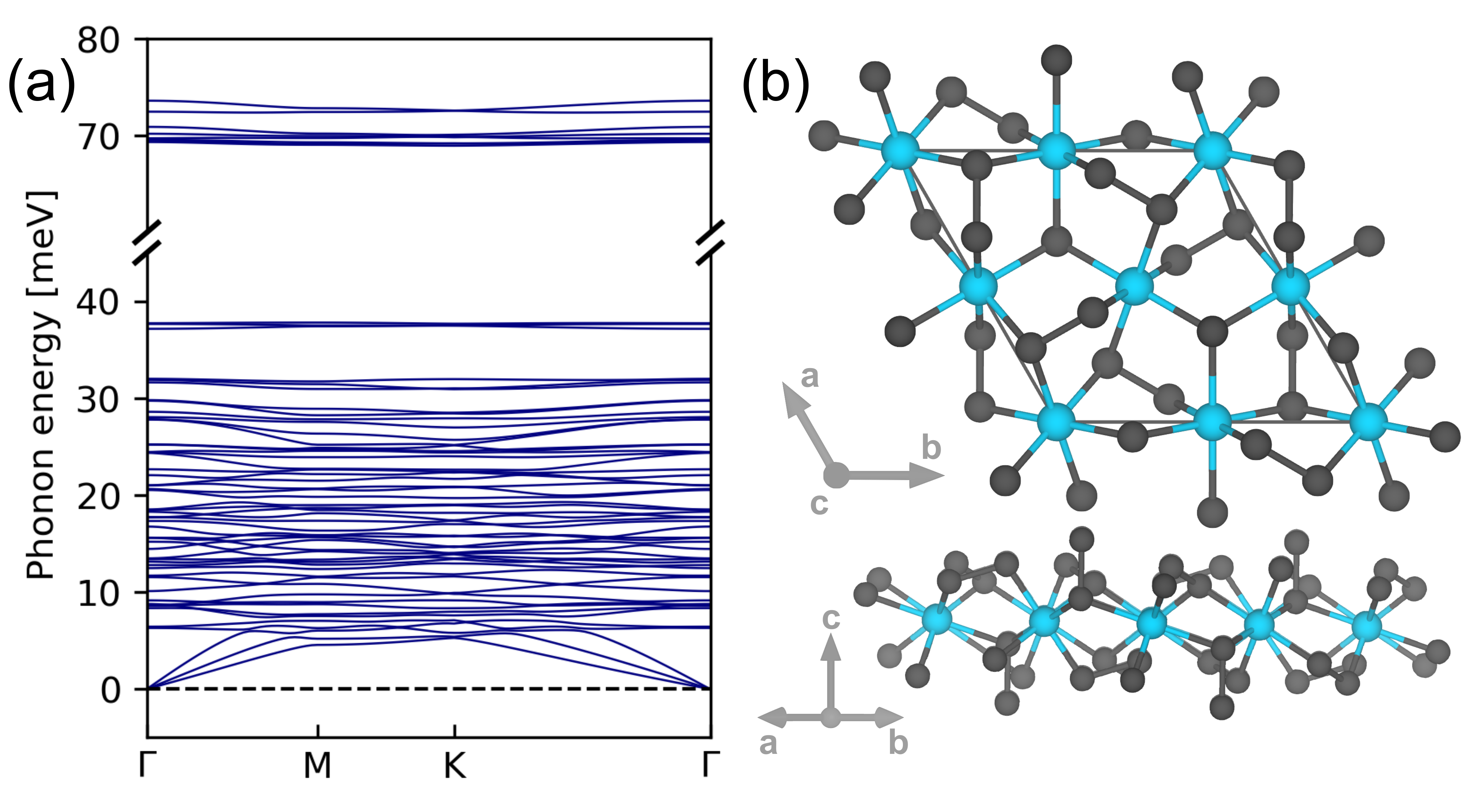}
\caption{\textbf{Structural Stability of Monolayer CuS\sub{2}}.
The predicted phonon dispersion of the monolayer structure in (a) 
shows no soft modes regardless 
of electronic temperature. Panel (b) depicts a top down and side perspective of the lattice 
structure, equivalent to the (111) close-packed plane cut from the original 
bulk cubic structure. The phonon results indicates an absence of CDW in the monolayer,
in accordance with recent experimental observations~\cite{liu_observation_2018}. 
This absence is consistent with the results of Fig.~\ref{fig:222supercell}, as 
the side-ways S\sub{2} motions could not be equivalently realized in the 2D limit.} 
\label{fig:monolayerComparison}
\end{figure}

Beyond our investigation of bulk CuS\sub{2}, we extend our work to include its monolayer,
which is formed as a quasi-layered hexagonal structure along the close-packed (111) plane as illustrated in Fig.~\ref{fig:monolayerComparison}(b). 
We predict a stable phonon dispersion for this structure, as shown in 
Fig.~\ref{fig:monolayerComparison}(a). Neither a Kohn anomaly nor soft 
phonon modes are observed for the same electronic temperature range, 
indicating the absence of CDW order in the monolayer limit.
This finding agrees with experimental results which saw no CDW signature in thin film CuS\sub{2} on a SrTiO\sub{3} substrate~\cite{liu_observation_2018}. 

This prediction raises questions as to why CDW order in bulk does not persist to the monolayer, in 
contrast to other layered TMDC
materials~\cite{ryu_persistent_2018,sakabe_direct_2017,singh_stable_2017,calandra_effect_2009}.
Understanding the difference between these two 
structures opens up a new avenue to understand CDW formation in the
bulk structure. We note that the monolayer cannot capture the side-ways S\sub{2} dimer motions seen in our calculated bulk distorted superstructure, and as a result, we find the stability of the monolayer supports our finding and predicted origin
of the potential CDW behavior in bulk CuS\sub{2}. However, a deeper understanding of the difference in bulk and monolayer structures prompts further study, particularly in attempts to increase the superconducting critical temperature in monolayer CuS\sub{2}. Therefore, we highlight the importance of this system as an ideal platform to study the coexistence of CDW order and superconductivity. 

While the structural phase diagram of CuS\sub{2} has remained a debate for years, we uncover critical information regarding its ability to manifest a CDW structure through theoretical and computational methods. 
We predict temperature-sensitive soft phonon modes that appear at low electronic temperatures in the experimentally reported structure of CuS\sub{2}, corresponding to a Kohn anomaly at the R point for an electronic temperature of approximately 100K. We then realize the existence of a stable distorted phase with a 2$a$ periodic lattice distortion and corresponding charge modulation at low electronic temperature. From this theoretical calculation, we establish a potential CDW transition and predict a crossover between the low and high symmetry structures at approximately 150K, in impressive agreement with what is reported in experiment. To further probe the electronic origins of the newly identified CDW structure, we observe a prominent band splitting near the Fermi energy at a wavevector half that of where the original instability occurred, similar to what is expected in the 1D perspective of CDW formation. Finally, we predict a stable monolayer with no CDW order signatures observed, in agreement with recent experiments, and note the potential difference in available dimer distortions as a possible reason for the lack of CDW in the monolayer.

Through this investigation of the phonon properties and electronic character of CuS\sub{2},
we report strong evidence of CDW in bulk CuS\sub{2}, suggesting
Fermi surface nesting as an explanation for its occurrence.
Additionally, because CuS\sub{2} lacks strong electronic correlations and simultaneously exhibits superconductivity, we identify this system as a promising platform to study the competition of superconductivity and CDW. In future work, we envision experimental studies of CDW order in bulk CuS\sub{2} (and absence thereof in monolayer) through angle resolved photoemission spectroscopy (ARPES and tr-ARPES), and low temperature scanning transmission electron microscopy (STEM).

\section*{Acknowledgements}
The authors acknowledge funding from the Defense Advanced Research Projects Agency
(DARPA) Defense Sciences Office (DSO) Driven and Nonequilibrium Quantum Systems program 
and the ONR grant on High-Tc Superconductivity at Oxide-Chalcogenide Interfaces (N00014-18-1-2691).
This research used resources of the National Energy Research Scientific 
Computing Center, a DOE Office of Science User Facility supported by 
the Office of Science of the U.S. Department of Energy under Contract 
No. DE-AC02-05CH11231, as well as resources at the Research Computing 
Group at Harvard University. JC recognizes the support of the DOE 
Computational Science Graduate Fellowship (CSGF) under grant 
DE-FG02-97ER25308.

\end{document}